\documentclass[journal,twocolumn]{IEEEtran}

\usepackage{epsfig,rotating,setspace,latexsym,amsmath,epsf,amssymb,amsfonts,bm,theorem,cite,algorithm,algorithmic,graphicx,epsf,authblk,epstopdf,color}

\newtheorem{theorem}{Theorem}
\newtheorem{lemma}{Lemma}
\newenvironment{Proof}[1]{\medskip\par\noindent{\bf Proof:\,}\,#1}{{\mbox{\,$\blacksquare$}\par}}

\allowdisplaybreaks

\begin{document}

\title{Optimal Policies for Wireless Networks with Energy Harvesting Transmitters and Receivers: Effects of Decoding Costs\thanks{Manuscript received March 29, 2015; revised June 26, 2015; accepted September 3, 2015. This work was supported by NSF Grants CNS 13-14733, CCF 14-22111 and CCF 14-22129, and was presented in part at the IEEE Global Conference on Signal and Information Processing (GlobalSIP), December 2014.}
\thanks{The authors are with the Department of Electrical and Computer Engineering, University of Maryland, College Park, MD 20742 (emails: {\it arafa@umd.edu}; {\it ulukus@umd.edu}).}
}

\author{Ahmed Arafa,~\IEEEmembership{Student Member,~IEEE} and~Sennur~Ulukus,~\IEEEmembership{Member,~IEEE}}

\maketitle

\begin{abstract}
We consider the effects of decoding costs in energy harvesting communication systems. In our setting, receivers, in addition to transmitters, rely solely on energy harvested from nature, and need to spend some energy in order to decode their intended packets. We model the decoding energy as an increasing convex function of the rate of the incoming data. In this setting, in addition to the traditional {\it energy causality} constraints at the transmitters, we have the {\it decoding causality} constraints at the receivers, where energy spent by the receiver for decoding cannot exceed its harvested energy. We first consider the point-to-point single-user problem where the goal is to maximize the total throughput by a given deadline subject to both energy and decoding causality constraints. We show that decoding costs at the receiver can be represented as {\it generalized data arrivals} at the transmitter, and thereby moving all system constraints to the transmitter side. Then, we consider several multi-user settings. We start with a two-hop network where the relay and the destination have decoding costs, and show that {\it separable} policies, where the transmitter's throughput is maximized irrespective of the relay's transmission energy profile, are optimal. Next, we consider the multiple access channel (MAC) and the broadcast channel (BC) where the transmitters and the receivers harvest energy from nature, and characterize the maximum departure region. In all multi-user settings considered, we decompose our problems into inner and outer problems. We solve the inner problems by exploiting the structure of the particular model, and solve the outer problems by water-filling algorithms.
\end{abstract}

\section{Introduction}

Energy harvesting communications offer the promise of energy self-sufficient, energy self-sustaining operation for wireless networks with significantly prolonged lifetimes. Energy harvesting communications have been considered mostly for energy harvesting \emph{transmitters}, e.g., \cite{jingP2P, kayaEmax, omurFade, ruiZhangEH, jingMAC, jingBC, elifBC, elifDataBC, omurBC, gunduz2Hop, ruiZhangRelay, erkipRelay, schoberRelay, letaiefRelay, varanBuffer, berkCoop, berkDiamond, kayaLoss, gunduzLoss, erkipCost, ruiZhangNonIdeal, orhan-broadband, omurHybrid, payaroMercury, letaiefTraining, vvvEH, ruiZhangOutage, varan-continuous, erkipDelay, aggarwalPmax}, with fewer works on energy harvesting \emph{receivers}, e.g., \cite{kayaRxEH, yatesRxEH1, yatesRxEH2, yatesRxEH3}. In this paper, we consider energy harvesting communications with both energy harvesting transmitters and receivers.

The energy harvested at the transmitters is used for data transmission according to a rate-power relationship, which is concave, monotone increasing in powers. The energy harvested at the receivers is used for decoding costs, which we assume to be convex, monotone increasing in the incoming rate \cite{grover, payaroRxEH, kayaRxEH, yatesRxEH1, rost-decPower, goldsmith-decPower}. The transmission energy costs and receiver decoding costs could be comparable, especially in short-distance communications, where high rates can be achieved with relatively low powers, and the decoding power could be dominant; see \cite{grover} and the references therein.

We model the energy needed for decoding at the receivers via \emph{decoding causality} constraints: the energy spent at the receiver for decoding cannot exceed the receiver's harvested energy. We already have the \emph{energy causality} constraints at the transmitter: the energy spent at the transmitter for transmitting data cannot exceed the transmitter's harvested energy. Therefore, for a given transmitter-receiver pair, transmitter powers need now to adapt to both energy harvested at the transmitter and at the receiver; the transmitter must only use powers, and therefore rates, that can be handled/decoded by the receiver.

The most closely related work to ours is \cite{kayaRxEH}, where the authors consider a general network with energy harvesting transmitters and receivers, and maximize a general utility function, subject to energy harvesting constraints at all terminals. Reference \cite{kayaRxEH} carries the effects of decoding costs to the objective function. If the objective function is no longer concave after this operation, it uses time-sharing to concavify it, leading to a convex optimization problem, which it then solves by using a generalized water-filling algorithm.

In this paper, we consider a similar problem with a specific utility function which is throughput, for specific network structures, with different decoding costs informed by network information theory. First, we consider the single-user channel, and observe that the decoding costs at the receiver can be interpreted as a \emph{gate keeper} at the front-end of the receiver that lets packets pass only if it has sufficient energy to decode. We show that we can carry this \emph{gate} effect to the transmitter as a \emph{generalized data arrival constraint}. Therefore, the setting with decoding costs at the receiver is equivalent to a setting with no decoding costs at the receiver, but with a (generalized) data arrival constraint at the transmitter \cite{jingP2P}. We also note that the energy harvesting component of the receiver can be separated as a \emph{virtual relay} between the transmitter and the receiver; and again, the problem can be viewed as a setting with no decoding costs at the receiver but with a \emph{virtual relay} with a (generalized) energy arrival constraint \cite{gunduz2Hop, ruiZhangRelay, erkipRelay, schoberRelay, letaiefRelay, varanBuffer}.

We then consider several multi-user settings. We begin with a decode-and-forward two-hop network, where the relay and the receiver both have decoding costs. This gives rise to {\it decode-and-forward causality} constraints at the relay in addition to decoding causality constraints at the receiver and energy causality constraints at the transmitter. We decompose the problem into inner and outer problems. In the inner problem, we fix the relay's decoding power strategy, and show that {\it separable} policies are optimal \cite{gunduz2Hop, ruiZhangRelay}. These are policies that maximize the throughput of the transmitter-relay link independent of maximizing the throughput of the relay-destination link. Thereby, we solve the inner problem as two single-user problems with decoding costs. In the outer problem, we find the best relay decoding strategy by a water-filling algorithm.

Next, we consider a two-user multiple access channel (MAC) with energy harvesting transmitters and receiver, and maximize the departure region. We consider two different decoding schemes: simultaneous decoding, and successive cancellation decoding \cite{elgamalKim}. Each scheme has a different decoding power consumption. For the simultaneous decoding scheme, we show that the boundary of the maximum departure region is achieved by solving a weighted sum rate maximization problem that can be decomposed into an inner and an outer problem. We solve the inner problem using the results of single-user fading problem \cite{omurFade}. The outer problem is then solved using a water-filling algorithm. In the successive cancellation decoding scheme, our problem formulation is non-convex. We then use a successive convex approximation technique that converges to a local optimal solution \cite{succCVXapprox, palomarGP}. The maximum departure region with successive cancellation decoding is larger than that with simultaneous decoding.

Finally, we characterize the maximum departure region of a two-user degraded broadcast channel (BC) with energy harvesting transmitter and receivers. With the transmitter employing superposition coding \cite{cover}, a corresponding decoding power consumption at the receivers is assumed. We again decompose the weighted sum rate maximization problem into an inner and outer problem. We show that the inner problem is equivalent to a classical single-user energy harvesting problem with a time-varying \emph{minimum power constraint}, for which we present an algorithm. We solve the outer problem using a water-filling algorithm similar to the outer problems of the two-hop network and the MAC with simultaneous decoding.

\section{Single-User Channel}\label{sec_p2p}

\begin{figure}[t]
\center
\includegraphics[scale=0.9]{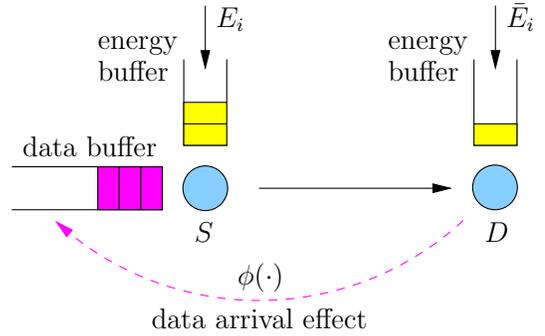}
\caption{Single-user channel with an energy harvesting transmitter and an energy harvesting receiver.}
\label{fig_p2p_sys}
\end{figure}

As shown in Fig.~\ref{fig_p2p_sys}, we have a transmitter and a receiver, both relying on energy harvested from nature. The time is slotted, and at the beginning of time slot $i\in \{1,\dots,N\}$, energies arrive at a given node ready to be used in the same slot or saved in a battery to be used in future slots. Let $\{E_i\}_{i=1}^N$ and $\{\bar{E}_i\}_{i=1}^N$ denote the energies harvested at each slot for the transmitter and the receiver, respectively, and let $\{p_i\}_{i=1}^N$ denote the transmitter's powers.

Without loss of generality, we assume that the time slot duration is normalized to one time unit. The physical layer is a Gaussian channel with zero-mean unit-variance noise. The objective is to maximize the total amount of data received \emph{and decoded} by the receiver by the deadline $N$. Our setting is \emph{offline} in the sense that all energy amounts are known prior to transmission.

The receiver must be able to decode the $k$th packet by the end of the $k$th slot. A transmitter transmitting at power $p_i$ in the $i$th time slot will send at a rate $g(p_i) \triangleq \frac{1}{2}\log_2\left(1+p_i\right)$, for which the receiver will spend $\phi(g(p_i))$ amount of power to decode, where $\phi$ is generally an increasing convex function \cite{grover, payaroRxEH, kayaRxEH, yatesRxEH1, rost-decPower, goldsmith-decPower}. In the sequel, we will also focus on the specific cases of linear and exponential functions, where $\phi(r)=ar + b$, with $a,b \geq 0$, and $\phi(r)=c2^{dr}+e$, with $c,d \geq 0$ and $c+e\geq0$. Continuing with a general convex increasing function $\phi$, we have the following decoding causality constraints for the receiver:
\begin{align}\label{eq_p2p_rx_battery}
\sum_{i=1}^k \phi(g(p_i)) \leq \sum_{i=1}^k \bar{E}_i , \quad k = 1,\dots,N
\end{align}
Therefore, the overall problem is formulated as:
\begin{align}\label{eq_opt_su_power}
\max_{{\bf p}} \quad &\sum_{i=1}^N g(p_i) \nonumber \\
\mbox{s.t.} \quad &\sum_{i=1}^k p_i \leq \sum_{i=1}^k E_i, \quad \forall k \nonumber \\
& \sum_{i=1}^k \phi(g(p_i)) \leq \sum_{i=1}^k \bar{E}_i, \quad \forall k
\end{align}
where $\mathbf{p}$ denotes the vector of powers. Note that the problem above in general is not a convex optimization problem as (\ref{eq_p2p_rx_battery}) in general is a non-convex constraint since $\phi$ is a convex function while $g$ is a concave function \cite{boyd}. Applying the change of variables $g(p_i)=r_i$, and defining $f\triangleq g^{-1}$ (note that $f$ is a convex function), we have
\begin{align}\label{eq_su_opt}
\max_{{\bf r}} \quad &\sum_{i=1}^N r_i \nonumber \\
\mbox{s.t.} \quad &\sum_{i=1}^k f(r_i) \leq \sum_{i=1}^k E_i, \quad \forall k \nonumber \\
& \sum_{i=1}^k \phi(r_i) \leq \sum_{i=1}^k \bar{E}_i, \quad \forall k
\end{align}
which is now a convex optimization problem \cite{boyd}.

We note that the constraints in (\ref{eq_p2p_rx_battery}), i.e., $\sum_{i=1}^k \phi(r_i) \leq \sum_{i=1}^k \bar{E}_i$, place upper bounds on the rates of the transmitter by every slot $k$. This resembles the problem addressed in \cite{jingP2P} with data packet arrivals during the communication session. In fact, when $\phi(r)=r$ and $\bar{E}_i=b_i$, where $b_i$ is the amount of data arriving in slot $i$, these are exactly the data arrival constraints in \cite{jingP2P}. A general convex $\phi$ generalizes this data arrival constraint. We characterize the solution of (\ref{eq_su_opt}) in the following three lemmas and the theorem. The proofs rely on the convexity of $f$ and $\phi$ generalizing the proof ideas in \cite{jingP2P}.

\begin{lemma}\label{thm_p2p_inc}
$\left\{ r_i^* \right\}$ is monotonically increasing.
\end{lemma}

\begin{Proof}
Assume that there exists a time slot $k$ such that $r_k^* > r_{k+1}^*$, and consider a new policy obtained by replacing both $r_k^*$ and $r_{k+1}^*$ by $\hat{r}_k = \hat{r}_{k+1} \triangleq \frac{r_k^*+r_{k+1}^*}{2}$, and observe that from the convexity of $f$ and $\phi$, we have
\begin{align}
f(\hat{r}_k) + f(\hat{r}_{k+1}) &\leq f(r_k^*) + f(r_{k+1}^*) \\
\phi(\hat{r}_k) + \phi(\hat{r}_{k+1}) &\leq \phi(r_k^*) + \phi(r_{k+1}^*)
\end{align}
In addition, since both $f$ and $\phi$ are monotonically increasing, we have $f\left(\hat{r}_k\right) \leq f\left(r_k^*\right)$, and $\phi\left(\hat{r}_k\right) \leq \phi\left(r_k^*\right)$. Therefore, the new policy is feasible, and can only save some energy either at the transmitter or at the receiver. This saved energy can be used to increase the rates in the upcoming time slots. Thus, the original policy cannot be optimal.
\end{Proof}

\begin{lemma}\label{thm_p2p_consume}
In the optimal policy, whenever the rate changes in a time slot, at least one of the following events occur: 1) the transmitter consumes all of its harvested energy in transmission, or 2) the receiver consumes all of its harvested energy in decoding, up to that time slot.
\end{lemma}

\begin{Proof}
Assume not, i.e., $r_k^* < r_{k+1}^*$ but both the transmitter and the receiver did not consume all their energies in the $k$th time slot. Then, we can always increase $r_k^*$ and decrease $r_{k+1}^*$ without conflicting the energy causality or the decoding causality constraints. By the convexity of $f$ and $\phi$, this modification would save some energy that can be used to increase the rates in the upcoming time slots. Therefore, the original policy cannot be optimal.
\end{Proof}

\begin{lemma}
In the optimal policy, by the end of the transmission period, at least one of the following events occur: 1) the transmitter's total power consumption in transmission is equal to its total harvested energy, or 2) the receiver's total power consumption in decoding is equal to its total harvested energy.
\end{lemma}

\begin{Proof}
Assume that both conditions are not met. Then, we can increase the rate in the last time slot until either the transmitter, or the receiver, consumes all of its energy. This is always feasible and strictly increases the rate.
\end{Proof}

\begin{theorem} \label{thm_p2p_sol}
Let $\psi \triangleq \phi^{-1}$. A policy is optimal iff it satisfies the following
\begin{align}\label{eq_p2p_sol1}
r_n=\min\left\{ g\left( \frac{\sum_{j=1}^{i_n} E_j - \sum_{j=1}^{i_{n-1}} f(r_j) }{i_n-i_{n-1}} \right), \right. \nonumber \\
\left. \psi\left( \frac{\sum_{j=1}^{i_n} \bar{E}_j - \sum_{j=1}^{i_{n-1}} \phi(r_j) }{i_n-i_{n-1}} \right) \right\}
\end{align}
where
\begin{align}\label{eq_p2p_sol2}
i_n=\arg \min_{i_{n-1} < i \leq N} \left\{ g\left( \frac{\sum_{j=1}^i E_j - \sum_{j=1}^{i_{n-1}} f(r_j) }{i-i_{n-1}} \right), \right. \nonumber \\
\left. \psi\left( \frac{\sum_{j=1}^i \bar{E}_j - \sum_{j=1}^{i_{n-1}} \phi(r_j) }{i-i_{n-1}} \right) \right\}
\end{align}
with $i_0=0$, and $n=1,\ldots,N$.
\end{theorem}

\begin{Proof}
First, we prove that the optimal policy satisfies (\ref{eq_p2p_sol1}) and (\ref{eq_p2p_sol2}). We show this by contradiction. Let us assume that the optimal policy, that satisfies the necessary lemmas above, is not given by (\ref{eq_p2p_sol1}) and (\ref{eq_p2p_sol2}) and achieves a higher throughput. In particular, let us assume that it coincides with the policy given by (\ref{eq_p2p_sol1}) and (\ref{eq_p2p_sol2}) for all rates $\{r_i\}_{i=1}^{n-1}$ but has a different value for $r_n$. Let us denote the points of rate increase of this policy by $\{i_k\}$. Thus, there must exist a time index $i^{\prime}>i_{n-1}$ such that
\begin{align}
r_n > \min\left\{ g\left( \frac{\sum_{j=1}^{i^{\prime}} E_j - \sum_{j=1}^{i_{n-1}} f(r_j) }{i^\prime-i_{n-1}} \right) , \right. \nonumber \\
\left. \psi\left( \frac{\sum_{j=1}^{i^{\prime}} \bar{E}_j - \sum_{j=1}^{i_{n-1}} \phi(r_j) }{i^\prime-i_{n-1}} \right) \right\}
\end{align}
and let us consider two different cases.

Assume that $i^{\prime}<i_n$. If the transmitter's energy is the bottleneck at $i^\prime$, then $r_n$ cannot be supported by the transmitter. On the other hand, if the receiver's energy is the bottleneck at $i^{\prime}$, then $r_n$ cannot be supported by the receiver. Hence, $r_n$ is not feasible in both cases. Now, assume that $i^{\prime}>i_n$. Then, there will exist a duration $\subseteq \left[ i_n +1 , i^{\prime} \right]$ where the rate has to decrease in order to satisfy feasibility. This violates the monotonicity property, and hence cannot be optimal.

Second, let us show sufficiency. We show this again by contradiction. Let us assume that the policy that satisfies (\ref{eq_p2p_sol1}) and (\ref{eq_p2p_sol2}) is not optimal. In particular, let us assume that there exists another policy $\left\{ r_i^{\prime} \right\}$ that coincides with it for all rates $\{r_i\}_{i=1}^{n-1}$ but has a different value for $r_n$. Since this new policy should have higher throughput, we have $r_n^{\prime}>r_n$. Now, assume $i_n^{\prime} > i_n$. Then, clearly $r_n^{\prime}$ is not feasible in the duration $\left[i_{n-1}+1,i_n\right]$. On the other hand, if $i_n^{\prime} < i_n$, then by the monotonicity property, all upcoming rates $\{r^{\prime}_i\}$ for $i > i_n^{\prime}$ can only be larger than $r_n^\prime$, which are all larger than $r_n$. This makes the new policy infeasible by the end of slot $i_n$ since $r_n$ consumes all feasible energy according to (\ref{eq_p2p_sol1}) and (\ref{eq_p2p_sol2}). Thus, the original policy is optimal.
\end{Proof}

Theorem~\ref{thm_p2p_sol} shows that decoding costs at the receiver are similar in effect to having a single-user channel with data arrivals during transmission and no decoding costs. This stems from the fact that the transmitter has to adapt its powers (and rates) in order to meet the decoding requirements at the receiver. Therefore, the receiver's harvested energies and the function $\phi$ control the amount of data the transmitter can send by any given point in time.

\begin{figure*}[t]
\center
\includegraphics[scale=0.9]{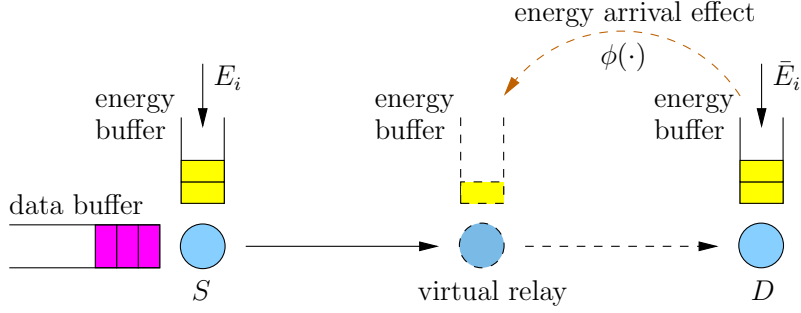}
\caption{Decoding costs viewed as a virtual relay.}
\label{fig_p2p_virtual}
\end{figure*}

Alternatively, we can slightly change the single-user problem (\ref{eq_su_opt}) by adding an extra variable $\bar{r}_i$ as follows
\begin{align}
\max_{{\bf r},\bar{{\bf r}}} \quad &\sum_{i=1}^N \bar{r}_i \nonumber \\
\mbox{s.t.} \quad &\sum_{i=1}^k f(r_i) \leq \sum_{i=1}^k E_i , \quad \forall k \nonumber \\
& \sum_{i=1}^k \phi(\bar{r}_i) \leq \sum_{i=1}^k \bar{E}_i , \quad \forall k \nonumber \\
&\bar{r}_i\leq r_i, \quad \forall i
\end{align}
This gives the same solution as we will always have $\bar{r}_i^*=r_i^*$ satisfied for all $i$. Therefore, as shown in Fig.~\ref{fig_p2p_virtual}, we can view the single-user setting with an energy harvesting receiver, as a two-hop setting with a \emph{virtual relay} between the transmitter and the receiver, with a non-energy harvesting receiver. To this end, we separate the decoding costs of the receiver, which are subject to energy harvesting constraints, as a relay which is subject to energy harvesting constraints in its transmissions, and consider the receiver as fully powered \cite{gunduz2Hop, erkipRelay, schoberRelay, ruiZhangRelay, letaiefRelay, varanBuffer}. The receiver will only receive data if the relay has sufficient energy to forward it. In addition, this energy harvesting virtual relay has no data buffer, thus, its incoming data rate equals its outgoing data rate. The rate through this relay is controlled by $\bar{E}_i$ and $\phi$. Thus, the decoding function $\phi$ puts a \emph{generalized energy arrival effect} to this virtual relay, in a similar way that it puts a \emph{generalized data arrival effect} to the transmitter through Theorem~\ref{thm_p2p_sol}, as shown in Fig.~\ref{fig_p2p_sys}.

It is worth mentioning that if we consider the special case where the receiver has no battery to store its energy, this will lead to the following decoding causality constraint
\begin{align}
\phi(g(p_i)) \leq \bar{E}_i, \quad  i = 1,\dots,N
\end{align}
which, in view of the generalized data arrival interpretation, can be modeled as a time-varying upper bound on the transmitter's power in each slot
\begin{align}
p_i^{\text{max}}\triangleq f\left( \psi\left( \bar{E}_i \right) \right)
\end{align}
where $\psi(\bar{E}_i)$ is the maximum transmission rate of a packet that $\bar{E}_i$ can handle at the decoder, and $p_i^{\text{max}}$ denotes its corresponding maximum transmit power. This problem has been considered in the general framework of \cite{palomarCVXsep}, and in \cite{aggarwalPmax} for the special case of a constant maximum power constraint. One solution for this problem is to apply a backward water-filling algorithm that starts from the last slot backwards, where at each slot directional water-filling \cite{omurFade} is applied only on slots whose maximum power constraint is not satisfied with equality. This might cause some wastage of water if the maximum power constraints are tighter than the transmitter's energy causality constraints, which depends primarily on how the function $\phi$ relates the transmitter's and the receiver's energies.

\section{Two-Hop Network}\label{sec_2h}

\begin{figure*}[t]
\center
\includegraphics[scale=0.9]{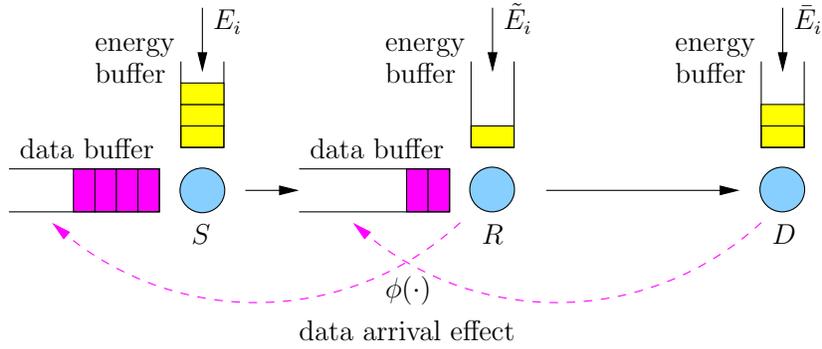}
\caption{Two-hop energy harvesting system with both relay and destination decoding costs.}
\label{fig_twohop}
\end{figure*}

We now consider a two-hop network consisting of a single source-destination pair communicating through a relay, as depicted in Fig.~\ref{fig_twohop}. The relay is full duplex, and it uses a decode-and-forward protocol. The relay has a data buffer to receive its incoming packets from the source. At the beginning of slot $i$, energies in the amounts of $E_i$, $\tilde{E}_i$, and $\bar{E}_i$ arrive at the source, relay, and destination, respectively. Unused energies can be saved in their respective batteries.

Let $r_i$ and $\tilde{r}_i$ be the rates of the source and the relay, respectively, in slot $i$. Our goal is to maximize the total amount of data received {\it and decoded} at the destination by the deadline $N$. We impose decoding costs on both the relay and the destination. The problem is formulated as:
\begin{align}\label{eq_2h_opt}
\max_{{\bf r}, \tilde{\bf r} } \quad &\sum_{i=1}^{N} \tilde{r}_i \nonumber \\
\mbox{s.t.} \quad &\sum_{i=1}^k f\left(r_i\right) \leq \sum_{i=1}^k E_i, \quad \forall k  \nonumber \\
&\sum_{i=1}^k \phi\left(r_i\right) + f\left(\tilde{r}_i\right)  \leq \sum_{i=1}^k \tilde{E}_i, \quad \forall k  \nonumber \\
&\sum_{i=1}^k \tilde{r}_i \leq \sum_{i=1}^k r_i, \quad \forall k  \nonumber \\
&\sum_{i=1}^k \phi\left(\tilde{r}_i\right)  \leq \sum_{i=1}^k \bar{E}_i, \quad \forall k
\end{align}
where the first constraint in (\ref{eq_2h_opt}) is the source transmission energy causality constraint, the second one is the relay decode-and-forward causality constraint, the third one is the data causality constraint at the relay, and the last one is the destination decoding causality constraint.

We first note that that if the relay did not have a data buffer, the source and the relay rates will need to be equal, i.e., $\tilde{r}_i=r_i$ for all $i$. In this case, the problem reduces to be a problem only in terms of the source rates, and could be solved by straightforward generalization of the single-user result in Theorem~\ref{thm_p2p_sol} considering three constraints instead of two. In a sense, this would be equivalent to taking the effects of decode-and-forward causality at the relay and decoding causality at the receiver back to the source as two different generalized data arrival effects. This can be further extended to multi-hop networks with relays having no data buffers by taking their constraint effects all the way back to the source.

In our setting, having a data buffer at the relay imposes non-obvious relationships among the source and the relay rates. To tackle this issue, we decompose the problem into inner and outer problems. In the inner problem, we solve for the source and relay rates after fixing a decoding power strategy for the relay node. By that we mean choosing the amounts of powers, $\{\delta_i\}_{i=1}^N$, the relay dedicates to decoding its incoming source packets. These amounts need to be feasible in the sense that $\sum_{i=1}^k \delta_i \leq \sum_{i=1}^k \tilde{E}_i$, $\forall k$. This decomposes the decode-and-forward causality constraint into the following two constraints:
\begin{align}
\sum_{i=1}^k \phi\left(r_i\right) \leq \sum_{i=1}^k \delta_i, \quad \sum_{i=1}^k f\left(\tilde{r}_i\right) \leq \sum_{i=1}^k \tilde{E}_i - \delta_i, \quad \forall k
\end{align}
In the next lemmas and theorem, we characterize the solution of the inner problem. The proofs of the lemmas are extensions of the ones presented in \cite{ruiZhangRelay} to the case of generalized data arrivals.

\begin{lemma}\label{thm_2h_inc}
There exists an optimal increasing source rate policy for the inner problem.
\end{lemma}

\begin{Proof}
Assume that there exists a time slot $k$ where $r_k>r_{k+1}$. We have two cases to consider. First, assume $\tilde{r}_k > \tilde{r}_{k+1}$. Let us define a new policy by replacing the $k$th and $k+1$st source and relay rates by  $r^\prime \triangleq \frac{r_k + r_{k+1}}{2}$, and $\tilde{r}^\prime \triangleq \frac{\tilde{r}_k + \tilde{r}_{k+1}}{2}$, respectively. By the convexity of $f$ and $\phi$, and linearity of the data causality constraint, the new policy is feasible, and can only save some energy at the source or the relay. This energy can be used in later slots to achieve higher rates.

Now, assume $\tilde{r}_k \leq \tilde{r}_{k+1}$. We argue that the data arrival causality constraint is satisfied with strict inequality at time slot $k$. For if it were equality, we need to have $\tilde{r}_k\geq r_k$ and $\tilde{r}_{k+1}\leq r_{k+1}$, which leads to $r_k \leq \tilde{r}_k \leq \tilde{r}_{k+1} \leq r_{k+1}$, an obvious contradiction. Now, we can find a small enough $\epsilon>0$, such that defining a new policy by replacing the $k$th and $k+1$st source rates by $r_k-\epsilon$ and $r_{k+1} + \epsilon$, respectively, we do not affect the relay rates. By the convexity of $f$ and $\phi$, the new policy is feasible, and can only save some energy at the source. This energy can be used in later slots to send more data to the relay, and hence, possibly increasing the relay rates, and the end-to-end throughput.
\end{Proof}

\begin{lemma}\label{thm_2h_sep}
The optimal increasing source rate policy for the inner problem $\{r_i^*\}$ is given by the single-user problem solution in (\ref{eq_p2p_sol1}) and (\ref{eq_p2p_sol2}), where the transmitter's and the receiver's energies are given by $\{E_i\}$ and $\{\delta_i\}$, respectively.
\end{lemma}

\begin{Proof}
Let us denote the single-user solution by $\{r_i^\prime\}$. Assume for contradiction that it is not optimal for the inner problem. In particular, let $\{r_i^*\}$ and $\{r_i^\prime\}$ be equal for $i=1,\dots,k-1$, and differ on the $k$th slot. We again have two cases to consider. First, assume $r_k^*>r_k^\prime$. In this case, since by Lemma~\ref{thm_2h_inc}, $\{r_i^*\}$ is increasing, by similar arguments as in the proof of Theorem~\ref{thm_p2p_sol}, the policy $\{r_i^*\}$ will eventually not satisfy the source's energy causality or the relay's decoding causality constraints, at some time slot $j\geq k$. Hence, it cannot be optimal.

Now, assume $r_k^*<r_k^\prime$. We argue that this shrinks the feasible set of the relay's rates. We show this by induction. By assumption of this case, it is true at time slot $k$, that we have $\sum_{i=1}^k r_i^*<\sum_{i=1}^k r_i^\prime$. Now, assume it is true that for some time slot $j>k$ we have $\sum_{i=1}^j r_i^*<\sum_{i=1}^j r_i^\prime$, and consider the $j+1$st time slot. If $r_{j+1}^*>r_{j+1}^\prime$, then we are back to the previous case where this cannot be feasible eventually. Therefore, the feasible set of the relay's rates shrinks at time slot $j+1$, and hence, shrinks all over $k,\dots,N$. Thus, this case cannot be optimal either.
\end{Proof}

Lemma~\ref{thm_2h_sep} states that the optimal source policy is separable \cite{gunduz2Hop, ruiZhangRelay} in the sense that the source maximizes its throughput to the relay irrespective of how the relay spends its transmission energy. This stems from the fact that the relay has an infinite data buffer to store its incoming source packets. Therefore, once we fix a decoding power strategy at the relay, we get separability. The following theorem, which is an extended version of Theorem~\ref{thm_p2p_sol}, gives the optimal relay rates for the inner problem. The proof is similar to that of Theorem~\ref{thm_p2p_sol} and is omitted for brevity.

\begin{theorem}
Given the optimal source rates $\{r_i^*\}$, the optimal relay rates for the inner problem is given by
\begin{align}\label{eq_2h_sol}
\tilde{r}_n^*&=\min\left\{ g\left( \frac{\sum_{j=1}^{i_n} \tilde{E}_j-\delta_j - \sum_{j=1}^{i_{n-1}} f(\tilde{r}_j^*) }{i_n-i_{n-1}} \right) , \right. \nonumber \\
&\left. \psi\left( \frac{\sum_{j=1}^{i_n} \bar{E}_j - \sum_{j=1}^{i_{n-1}} \phi(\tilde{r}_j^*) }{i_n-i_{n-1}} \right) , \frac{\sum_{j=1}^{i_n} r_j^* - \sum_{j=1}^{i_{n-1}} r_j^*}{i_n - i_{n-1}} \right\}
\end{align}
where $i_n$ is the $\arg \min$ of the expression in (\ref{eq_2h_sol}) as in (\ref{eq_p2p_sol1})-(\ref{eq_p2p_sol2}), and $i_0=0$.
\end{theorem}

Denoting the solution of the inner problem by $R(\bm{\delta})$, we now find the optimal relay decoding strategy $\{\delta_i^*\}$ by solving the following outer problem:
\begin{align}
\max_{\bm{\delta}} \quad &R(\bm{\delta}) \nonumber \\
\mbox{s.t.} \quad &\sum_{i=1}^k \delta_i \leq \sum_{i=1}^k \tilde{E}_i, \quad \forall k
\end{align}
We have the following lemma regarding the outer problem.

\begin{lemma}
$R(\bm{\delta})$ is a concave function.
\end{lemma}

\begin{Proof}
Consider two decoding power strategies $\bm{\delta}_1$, $\bm{\delta}_2$, and let $\{{\bf r}_1,\tilde{\bf r}_1\}$, $\{{\bf r}_2,\tilde{\bf r}_2\}$ be their corresponding source and relay optimal inner problem rates, respectively. Let $\bm{\delta}_\theta\triangleq \theta\bm{\delta}_1+(1-\theta)\bm{\delta}_2$, for some $0\leq\theta\leq1$, and consider the rate policy defined by ${\bf r}_\theta\triangleq\theta{\bf r}_1+(1-\theta){\bf r}_2$, and $\tilde{\bf r}_\theta\triangleq\theta\tilde{\bf r}_1+(1-\theta)\tilde{\bf r}_2$, for the source, and the relay, respectively. By the convexity of $f$ and $\phi$, the policy $\{{\bf r}_\theta,\tilde{\bf r}_\theta\}$ is feasible for the decoding strategy $\bm{\delta}_\theta$. Therefore, we have
\begin{align}
R(\bm{\delta}_\theta) \geq \sum_{i=1}^N \tilde{r}_{\theta i} = \theta R(\bm{\delta}_1) + (1-\theta) R(\bm{\delta}_2)
\end{align}
proving the concavity of $R(\bm{\delta})$.
\end{Proof}

Therefore, the outer problem is a convex optimization problem \cite{boyd}. We propose a water-filling algorithm to solve the outer problem \cite{berkDiamond}. We first note that $R(\bm{\delta})$ does not possess any monotonicity properties in the feasible region. For instance, $R(\tilde{\bf E})=R({\bf 0})=0$, while $R(\bm{\delta})$ is strictly positive for some $\bm{\delta}$ in between. Thus, at the optimal relay decoding power strategy, not all the relay's decoding energy will be exhausted. To this end, we add an extra $N+1$st slot where we can possibly discard some energy. We start by filling up each slot by its corresponding energy/water level and we leave the extra $N+1$st slot initially empty. Meters are put in between bins to measure the amount of water passing. We let water flow to the right only if this increases the objective function. After each iteration, water can be called back if this increases the objective function. All the amount of water that is in the extra slot is eventually discarded, but may be called back also during the iterations. Since with each water flow the objective function monotonically increases, problem feasibility is maintained throughout the process, and due to the convexity of the problem, the algorithm converges to the optimal solution.

\section{Multiple Access Channel}

\begin{figure}[t]
\center
\includegraphics[scale=0.9]{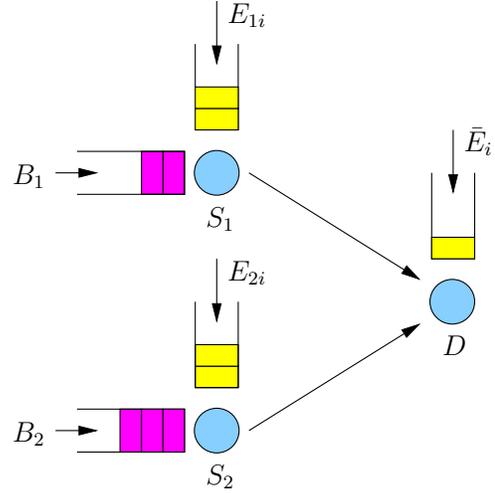}
\caption{Two-user MAC with energy harvesting transmitters and receiver.}
\label{fig_mac_sys}
\end{figure}

We now consider a two-user Gaussian MAC as shown in Fig.~\ref{fig_mac_sys}. The two transmitters harvest energy in amounts $\left\{ E_{1i} \right\}_{i=1}^N$ and $\left\{ E_{2i} \right\}_{i=1}^N$, respectively, and the receiver harvests energy in amounts $\left\{\bar{E}_i\right\}_{i=1}^N$. The receiver noise is with zero-mean and unit-variance. The capacity region for this channel is given by \cite{cover}:
\begin{align}
r_1 &\leq g(p_1) \nonumber \\
r_2 &\leq g(p_2) \nonumber \\
r_1+r_2 &\leq g(p_1+p_2)
\end{align}
where $p_1$ and $p_2$ are the powers used by the first and the second transmitter, respectively.

In addition to the usual energy harvesting causality constraints on the transmitters \cite{jingMAC}, we impose a receiver decoding cost. We note that there can be different ways to impose this constraint depending on how the receiver employs the decoding procedure. In the next two sub-sections, we consider two kinds of decoding procedures, namely, simultaneous decoding, and successive decoding \cite{cover, elgamalKim}. Changing the decoding model affects the optimal power allocation for both users so as to adapt to how the receiver spends its power.

\subsection{Simultaneous Decoding}

In this case, the two transmitters can only send at rates whose sum can be decoded at the receiver. A power control policy $\{p_{1i},p_{2i}\}_{i=1}^N$ is feasible if the following are satisfied:
\begin{align}\label{eq_mac_feasibility}
\sum_{i=1}^k p_{1i} &\leq \sum_{i=1}^k E_{1i}, \quad \forall k \nonumber \\
\sum_{i=1}^k p_{2i} &\leq \sum_{i=1}^k E_{2i}, \quad \forall k \nonumber \\
\sum_{i=1}^k \phi\left(g\left(p_{1i}+p_{2i}\right)\right) &\leq \sum_{i=1}^k \bar{E}_i, \quad \forall k
\end{align}
From here on, we assume a specific structure for the decoding function $\phi$ for mathematical tractability and ease of presentation. In particular, we assume that it is exponential with parameters $c=1$, $d=2$ and $e=-1$, i.e., $\phi(r)=g^{-1}(r)=2^{2r}-1$. Let $B_j$ denote the total departed bits from the $j$th user by time slot $N$. Our aim is to characterize the \emph{maximum departure region}, $\mathcal{D}(N)$, which is the region of $(B_1,B_2)$ the transmitters can depart by time slot $N$, through a feasible policy. The following lemmas characterize this region \cite{jingMAC}.

\begin{lemma}
The maximum departure region, $\mathcal{D}(N)$, is the union of all $(B_1,B_2)$, over all feasible policies $\{p_{1i},p_{2i}\}_{i=1}^N$, where for any fixed power policy, $(B_1,B_2)$ satisfy
\begin{align}
B_1 &\leq \sum_{i=1}^{N} g(p_{1i}) \nonumber \\
B_2 &\leq \sum_{i=1}^{N} g(p_{2i}) \nonumber \\
B_1 + B_2 &\leq \sum_{i=1}^{N} g(p_{1i}+p_{2i})
\end{align}
\end{lemma}

\begin{lemma}
$\mathcal{D}(N)$ is a convex region.
\end{lemma}

\begin{figure}[t]
\center
\includegraphics[scale=0.7]{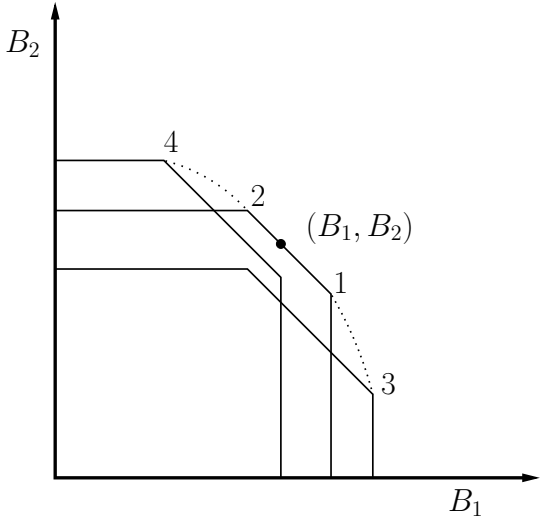}
\caption{Departure region of a two-user MAC.}
\label{fig_mac_region}
\end{figure}

Each point on the boundary of $\mathcal{D}(N)$, see Fig. \ref{fig_mac_region}, can be characterized by solving a weighted sum rate maximization problem subject to feasibility conditions (\ref{eq_mac_feasibility}). Let $\mu_1$ and $\mu_2$ be the non-negative weights for the first and the second user rates, respectively. Assuming without loss of generality that $\mu_1 > \mu_2$, and defining $\mu \triangleq \frac{\mu_2}{\mu_1-\mu_2}$, we then need to solve the following optimization problem:
\begin{align}\label{eq_mac_caseC}
\max_{{\bf p}_1, {\bf p}_2 } \quad &\sum_{i=1}^N g(p_{1i}) + \mu\sum_{i=1}^N g(p_{1i} + p_{2i}) \nonumber \\
\mbox{s.t.} \quad &\sum_{i=1}^k p_{1i} \leq \sum_{i=1}^k E_{1i}, \quad \forall k \nonumber \\
& \sum_{i=1}^k p_{2i} \leq \sum_{i=1}^k E_{2i}, \quad \forall k \nonumber \\
& \sum_{i=1}^k p_{1i} + p_{2i} \leq \sum_{i=1}^k \bar{E}_i, \quad \forall k
\end{align}
We note that the above problem resembles the one formulated in \cite{berkDiamond} for a diamond channel with energy cooperation. First, we state a necessary condition of optimality for the above problem.

\begin{lemma}
In the optimal solution for (\ref{eq_mac_caseC}), by the end of the transmission period, at least one of the following occur: 1) both transmitters consume all of their harvested energies in transmission, or 2) the receiver consumes all of its harvested energy in decoding.
\end{lemma}

\begin{Proof}
Assume without loss of generality that transmitter 1 does not consume all of its energies in transmission, and that the receiver also does not consume all of its energies in decoding. Then, we can always increase the value of $p_{1N}$ until either transmitter 1 or the receiver consume their energies. This strictly increases the objective function.
\end{Proof}

We decompose the optimization problem (\ref{eq_mac_caseC}) into two nested problems. First, we solve for ${\bf p}_2$ in terms of ${\bf p}_1$, and then solve for ${\bf p}_1$. Let us define the following inner problem:
\begin{align}\label{eq_mac_inner_prb}
G({\bf p}_1) \triangleq \max_{{\bf p}_2 } \quad & \sum_{i=1}^N g(p_{1i}+p_{2i}) \nonumber \\
\mbox{s.t.} \quad &\sum_{i=1}^k p_{2i} \leq \sum_{i=1}^k Q_i , \quad \forall k
\end{align}
where the modified energy levels $Q_i$ are defined as follows:
\begin{align}\label{eq_mac_Q}
&Q_i=M_i-M_{i-1}, \nonumber \\
&M_i = \min\left\{ \sum_{j=1}^i E_{2j}, \sum_{j=1}^i \bar{E}_j - p_{1j} \right\}, \quad M_0=0
\end{align}
Then, we have the following lemma.

\begin{lemma}\label{thm_inner_prb_ccv}
$G({\bf p}_1)$ is a decreasing concave function in ${\bf p}_1$.
\end{lemma}

\begin{Proof}
$G$ is a decreasing function of ${\bf p}_1$ since the feasible set shrinks with ${\bf p}_1$. To show concavity, let us choose two points ${\bf p}_1^{(1)}$ and ${\bf p}_1^{(2)}$, and take their convex combination ${\bf p}_1^{\theta}=\theta {\bf p}_1^{(1)} + (1-\theta) {\bf p}_1^{(2)}$ for some $0 \leq \theta \leq 1$. Let ${\bf p}_2^{(1)}$ and ${\bf p}_2^{(2)}$ denote the solutions of the inner problem (\ref{eq_mac_inner_prb}) at ${\bf p}_1^{(1)}$ and ${\bf p}_1^{(2)}$, respectively. Now, let ${\bf p}_2^{\theta} \triangleq \theta{\bf p}_2^{(1)} + (1-\theta){\bf p}_2^{(2)}$, and observe that, from the linearity of the constraint set, ${\bf p}_2^{\theta}$ is feasible with respect to ${\bf p}_1^{\theta}$. Therefore, we have
\begin{align}
G\left({\bf p}_1^{\theta} \right) &\geq \sum_{i=1}^N g\left(p_{1i}^{\theta}+p_{2i}^{\theta}\right) \nonumber \\
&\geq \sum_{i=1}^N \theta g\left(p_{1i}^{(1)}+p_{2i}^{(1)}\right) + (1-\theta) g\left(p_{1i}^{(2)}+p_{2i}^{(2)}\right) \nonumber \\
&= \theta G\left({\bf p}_1^{(1)} \right) + (1-\theta) G\left({\bf p}_1^{(2)} \right)
\end{align}
where the second inequality follows from the concavity of $g$.
\end{Proof}

We observe that the inner problem (\ref{eq_mac_inner_prb}) is a single-user energy harvesting maximization problem with fading, whose solution is via directional water-filling of $\{Q_i\}_{i=1}^N$ over the inverse of the fading levels $\{1+p_{1i}\}_{i=1}^N$ as presented in \cite{omurFade}. Next, we solve the outer problem given by:
\begin{align}
\max_{{\bf p}_1 } \quad &\mu G\left({\bf p}_1\right) + \sum_{i=1}^N g(p_{1i}) \nonumber \\
\mbox{s.t.} \quad &\sum_{i=1}^k p_{1i} \leq \sum_{i=1}^k T_i , \quad \forall k
\end{align}
where we define the water levels $T_i=L_i - L_{i-1}$, with $L_i = \min\left\{ \sum_{j=1}^i E_{1j}, \sum_{j=1}^i \bar{E}_j\right\}$, and $L_0=0$. The minimum is added to ensure the feasibility of the inner problem. Note that, by Lemma~\ref{thm_inner_prb_ccv}, the outer problem is a convex optimization problem \cite{boyd}. We first note that at the optimal policy, first user's modified energies $\{T_i\}$ need not be fully utilized by the end of transmission. This is because the objective function is not increasing in ${\bf p}_1$. To this end, we use the iterative water-filling algorithm for the outer problem proposed in Section~\ref{sec_2h} to solve this outer problem. Since the problem is convex, iterations converge to the optimal solution.

Note that the above formulation obtains the dotted points in the curved portion of the departure region in Fig.~\ref{fig_mac_region}. Specific points in the departure region, e.g., points 1 and 3 in Fig.~\ref{fig_mac_region}, can be found by specific schemes \cite{arafaGlobalSIP}, by solving the problem for the cases $\mu_1=\mu_2$ and $\mu_1\mu_2=0$.

\subsection{Successive Cancelation Decoding}

We now let the receiver employ successive decoding, where it aims at decoding the corner points, and then uses time sharing if necessary to achieve the desired rate pair \cite{cover, elgamalKim}. For instance, if the system is operating at its lower corner point, then the receiver first decodes the message of the second user, by treating the first user's signal as noise, then decodes the message of the first user, after subtracting the second user's signal from its received signal. For $\mu_1 > \mu_2$, we are always at a lower corner point at every time slot, and therefore the weighted sum rate maximization problem can be formulated as:
\begin{align}\label{eq_mac_succ_powers}
\max_{{\bf p}_1, {\bf p}_2 } \quad &\mu_1\sum_{i=1}^N g(p_{1i}) + \mu_2\sum_{i=1}^N g \left( \frac{p_{2i}}{1+p_{1i}} \right) \nonumber \\
\mbox{s.t.} \quad &\sum_{i=1}^k p_{1i} \leq \sum_{i=1}^k E_{1i}, \quad \forall k \nonumber \\
&\sum_{i=1}^k p_{2i} \leq \sum_{i=1}^k E_{2i}, \quad \forall k \nonumber \\
& \sum_{i=1}^k p_{1i} + \frac{p_{2i}}{1+p_{1i}} \leq \sum_{i=1}^k \bar{E}_i, \quad \forall k
\end{align}
where the last inequality comes from the fact that the receiver is decoding the second user's message first by treating the first user's signal as noise, and thereby spends $\phi\left(g\left(\frac{p_{2i}}{1+p_{1i}}\right)\right)$ amount of energy to decode this message, and then spends $\phi\left(g\left(p_{1i}\right)\right)$ amount of energy to decode the first user's message after subtracting the second user's signal.

Observe that the last constraint, i.e., the decoding causality constraint, is non-convex. Therefore, one might need to invoke the time-sharing principle in order to fully characterize the boundary of the maximum departure region. In terms of the rates the problem can be written as:
\begin{align}
\max_{{\bf r}_1, {\bf r}_2 } \quad &\mu_1\sum_{i=1}^N r_{1i} + \mu_2\sum_{i=1}^N r_{2i} \nonumber \\
\mbox{s.t.} \quad &\sum_{i=1}^k 2^{2r_{1i}} - 1 \leq \sum_{i=1}^k E_{1i}, \quad \forall k \nonumber \\
&\sum_{i=1}^k 2^{2r_{1i}}\left(2^{2r_{2i}} - 1\right) \leq \sum_{i=1}^k E_{2i}, \quad \forall k \nonumber \\
& \sum_{i=1}^k 2^{2r_{1i}} + 2^{2r_{2i}} - 2 \leq \sum_{i=1}^k \bar{E}_i, \quad \forall k
\end{align}
which is a non-convex problem due to the second user's energy causality constraint. In fact, the above problem is a signomial program, a generalized form of a geometric program, where posynomials can have negative coefficients \cite{boyd}. Next, we use the idea of successive convex approximation \cite{succCVXapprox} to provide an algorithm that converges to a local optimal solution.

By applying the change of variables $x_{ji} \triangleq 2^{2r_{ji}}-1$, $j=1,2$, and some algebraic manipulations:
\begin{align}
\min_{{\bf x}_1, {\bf x}_2 , {\bf t}_1 , {\bf t}_2 } \quad &\sum_{i=1}^N t_{1i}^{-\mu_1} t_{2i}^{-\mu_2} \nonumber \\
\mbox{s.t.} \qquad &\sum_{i=1}^k x_{1i} \leq \sum_{i=1}^k E_{1i}, \quad \forall k \nonumber \\
&\sum_{i=1}^k \left(1+x_{1i}\right)x_{2i} \leq \sum_{i=1}^k E_{2i}, \quad \forall k \nonumber \\
& \sum_{i=1}^k x_{1i} + x_{2i} \leq \sum_{i=1}^k \bar{E}_i, \quad \forall k \nonumber \\
&t_{1i} \leq 1+x_{1i}, \quad \forall i \nonumber \\
&t_{2i} \leq 1+x_{2i}, \quad \forall i
\end{align}

Now, the problem looks very similar to a geometric program except for the last two sets of constraints. These constraints are written in the form of a monomial less than a posynomial, which will not allow us to write the problem in convex form by the usual geometric programming transformations \cite{boyd}. We will follow an approach introduced in \cite{palomarGP} in order to iteratively approximate the posynomials on the right hand side by monomials, and thereby reaching a geometric program that can be efficiently solved \cite{boyd}. Approximations should be chosen carefully such that iterations converge to a local optimum solution of the original problem \cite{succCVXapprox}. Towards that, we use the arithmetic-geometric mean inequality to write:
\begin{align}
1+x \geq \left(\frac{1}{\alpha}\right)^\alpha \left( \frac{x}{1-\alpha} \right)^{1-\alpha} \triangleq u(x;\alpha)
\end{align}
which holds for $0 \leq \alpha \leq 1$. In particular, equality holds at a point $x_k \geq 0$ if we choose $\alpha=\frac{1}{1+x_k}$. Therefore, the monomial function $u(x;\alpha_k)$ approximates the posynomial function $1+x$ at $x=x_k$. Substituting this approximation, we obtain that at the $k+1$st iteration, we need to solve the following geometric program:
\begin{align}\label{eq_mac_succ_cvxApprox}
\min_{{\bf x}_1, {\bf x}_2 , {\bf t}_1 , {\bf t}_2 } \quad &\sum_{i=1}^N t_{1i}^{-\mu_1} t_{2i}^{-\mu_2} \nonumber \\
\mbox{s.t.} \qquad  &\sum_{i=1}^k x_{1i} \leq \sum_{i=1}^k E_{1i}, \quad \forall k \nonumber \\
&\sum_{i=1}^k \left(1+x_{1i}\right)x_{2i} \leq \sum_{i=1}^k E_{2i}, \quad \forall k \nonumber \\
& \sum_{i=1}^k x_{1i} + x_{2i} \leq \sum_{i=1}^k \bar{E}_i, \quad \forall k \nonumber \\
&\frac{t_{1i}}{u\left( x_{1i} ; \alpha_{1i}^{(k)} \right)} \leq 1, \quad \forall i \nonumber \\
&\frac{t_{2i}}{u\left( x_{2i} ; \alpha_{2i}^{(k)} \right)} \leq 1, \quad \forall i
\end{align}
where $\alpha_{ji}^{(k)} \triangleq \frac{1}{1+x_{ji}^{(k)}}$, $j=1,2$, and $x_{ji}^{(k)}$ is the solution of the $k$th iteration. We pick an initial feasible point $\left( {\bf x}_1^{(0)} , {\bf x}_2^{(0)} \right)$ and run the iterations. The choice of the approximating monomial function $u$ satisfies the conditions of convergence stated in \cite{succCVXapprox}, and therefore, the iterative solution of problem (\ref{eq_mac_succ_cvxApprox}) converges to a point $\left( {\bf x}_1^* , {\bf x}_2^* \right)$ that is local optimal for problem (\ref{eq_mac_succ_powers}). Finally, we get the original power allocations by substituting $p_{1i}^*=x_{1i}^*$, and $p_{2i}^*=\left( x_{1i}^* + 1 \right) x_{2i}^*$.

\section{Broadcast Channel}

\begin{figure}[t]
\center
\includegraphics[scale=0.9]{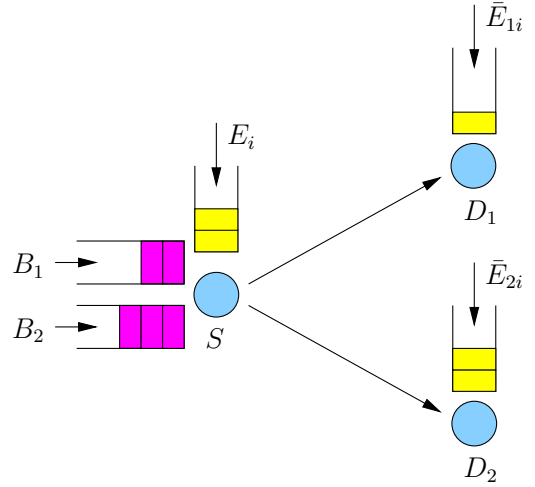}
\caption{Two-user BC with energy harvesting transmitter and receivers.}
\label{fig_bc_sys}
\end{figure}

We now consider a two-user Gaussian BC with energy harvesting transmitter and receivers as shown in Fig.~\ref{fig_bc_sys}. Energies arrive in amounts $E_i,$ $\bar{E}_{1i},$ and $\bar{E}_{2i},$ at the transmitter, and the receivers 1 and 2, respectively. By superposition coding \cite{cover}, the weaker user is required to decode its message while treating the stronger user's interference as noise. While the stronger user is required to decode both messages successively by first decoding the weaker user's message, and then subtracting it to decode its own. The receiver noises have variances 1 and $\sigma^2 > 1$.

Under a total transmit power $P$, the capacity region of the Gaussian BC is \cite{cover}:
\begin{align}
r_1 \leq \frac{1}{2} \log_2 \left( 1 + \alpha P \right ), ~ r_2 \leq \frac{1}{2} \log_2 \left( 1 + \frac{(1-\alpha) P}{\alpha P + \sigma^2} \right )
\end{align}
working on the boundary of the capacity region we have:
\begin{align}
P = \left( \sigma^2 -1 \right) 2^{2r_2} + 2^{2\left(r_1+r_2\right)} - \sigma^2 \triangleq F\left( r_1,r_2 \right)
\end{align}
where $F(r_1,r_2)$ is the minimum power needed by the transmitter to achieve rates $r_1$ and $r_2$. Note that $F$ is an increasing convex function of both rates.

As in the MAC case, the goal here is to characterize the maximum departure region:
\begin{align}\label{eq_opt_bc}
\max_{{\bf r_1 , r_2} } \quad &\mu_1\sum_{i=1}^N r_{1i} + \mu_2\sum_{i=1}^N r_{2i}  \nonumber \\
\mbox{s.t.} \quad &\sum_{i=1}^k F\left( r_{1i} , r_{2i} \right) \leq \sum_{i=1}^k E_i, \quad \forall k \nonumber \\
& \sum_{i=1}^k \phi(r_{1i}+r_{2i}) \leq \sum_{i=1}^k \bar{E}_{1i}, \quad \forall k \nonumber \\
& \sum_{i=1}^k \phi(r_{2i}) \leq \sum_{i=1}^k \bar{E}_{2i}, \quad \forall k
\end{align}
where the first constraint in (\ref{eq_opt_bc}) is the source transmission energy causality constraint, and second and third constraints are the decoding causality constraints at the stronger and weaker receivers, respectively. Here also, we take the decoding cost function $\phi$ to be $\phi(r)=2^{2r}-1$.

By virtue of superposition coding, we see that, in the optimization problem in (\ref{eq_opt_bc}), the decoding causality constraint of the stronger user is a function of both rates intended for the two users, as it is required to decode both messages. While the decoding causality constraint for the weaker user is a function of its own rate only. By the convexity of $F$ and $\phi$, the maximum departure region is convex, and thus the weighted sum rate maximization in (\ref{eq_opt_bc}) is sufficient to characterize its boundary \cite{jingBC}. In addition, the  optimization problem in (\ref{eq_opt_bc}) is convex \cite{boyd}.

We note that a related problem has been considered in \cite{elifDataBC}, where the authors characterized transmission completion time minimization policies for a BC setting with data arrivals during transmission. There, the solution is found by sequentially solving an equivalent energy consumption minimization problem until convergence. Their solution is primarily dependent on Newton's method \cite{boyd}. Some structural insights are also presented about the optimal solution. In our setting, we consider the case with receiver side decoding costs, and generalize the data arrivals concept by considering the convex function $\phi$. In addition, our formulation imposes further interactions between the strong and the weak user's data, by allowing a constraint (strong user's) that is put on the sum of both rates, instead of on individual rates.

We characterize the solution of the problem according to the relation between $\mu_1$ and $\mu_2$ as follows. If $\mu_1 \geq \mu_2$, then due to the degradedness of the second user, it is optimal to put all power into the first user's message. This way, the problem reduces to a single-user problem:
\begin{align}
\max_{{\bf r_1} } \quad &\sum_{i=1}^N r_{1i}  \nonumber \\
\mbox{s.t.} \quad &\sum_{i=1}^k 2^{2r_{1i}} - 1 \leq \sum_{i=1}^k W_i, \quad \forall k
\end{align}
where the modified energy levels $\{W_i\}$ are defined as follows:
\begin{align}
&W_i=L_i-L_{i-1}, \nonumber \\
&L_i = \min\left\{ \sum_{j=1}^i E_j, \sum_{j=1}^i \bar{E}_{1j} \right\}, \quad L_0=0
\end{align}

On the other hand, if $\mu_1 < \mu_2$, then we need to investigate the necessary KKT optimality conditions \cite{boyd}. We write the Lagrangian for the problem (\ref{eq_opt_bc}) as follows:
\begin{align}
\mathcal{L}=&-\mu_1\sum_{i=1}^N r_{1i} - \mu_2\sum_{i=1}^N r_{2i} \nonumber \\
&+ \sum_{i=1}^N \lambda_i \left(\sum_{j=1}^i \left(\sigma^2-1\right)2^{2r_{2j}} + 2^{2(r_{1j}+r_{2j})} - \sigma^2 - E_j\right) \nonumber \\
&+ \sum_{i=1}^N \nu_{1i} \left(\sum_{j=1}^i 2^{2(r_{1j}+r_{2j})} - 1 - \bar{E}_{1j}\right) \nonumber \\
&+ \sum_{i=1}^N \nu_{2i} \left(\sum_{j=1}^i 2^{2r_{2j}} - 1 - \bar{E}_{2j}\right) \nonumber \\
&- \sum_{i=1}^N \eta_{1i}r_{1i} - \sum_{i=1}^N \eta_{2i}r_{2i}
\end{align}
Taking the derivative with respect to $r_{1i}$ and $r_{2i}$ and equating to zero, we obtain:
\begin{align}
2^{2(r_{1i}+r_{2i})}&=\frac{\mu_1+\eta_{1i}}{\sum_{j=i}^N \lambda_j+\nu_{1j}} \label{eq_bc_sumrate} \\
2^{2r_{2i}}&=\frac{\mu_2-\mu_1+\eta_{2i}-\eta_{1i}}{\sum_{j=i}^N (\sigma^2-1)\lambda_j+\nu_{2j}} \label{eq_bc_weakrate}
\end{align}
along with the complementary slackness conditions:
\begin{align}
&\lambda_i \left(\sum_{j=1}^i \left(\sigma^2-1\right)2^{2r_{2j}} + 2^{2(r_{1j}+r_{2j})} - \sigma^2 - E_j\right)=0, \quad \forall i \nonumber \\
&\nu_{1i} \left(\sum_{j=1}^i 2^{2(r_{1j}+r_{2j})} - 1 - \bar{E}_{1j}\right)=0, \quad \forall i \nonumber \\
&\nu_{2i} \left(\sum_{j=1}^i 2^{2r_{2j}} - 1 - \bar{E}_{2j}\right)=0, \quad \forall i \nonumber \\
&\eta_{1i}r_{1i}=0, \quad \eta_{2i}r_{2i}=0, \quad \forall i
\end{align}
From here, we state the following lemmas

\begin{lemma}\label{thm_bc_sumrate}
The sum rate $\{r_{1i}^*+r_{2i}^*\}$ is monotonically increasing.
\end{lemma}

\begin{Proof}
We prove this by contradiction. Assume that there exists some time slot $k$ such that $r_{1k}+r_{2k}>r_{1(k+1)}+r_{2(k+1)}$. From (\ref{eq_bc_sumrate}), since the denominator cannot increase, the numerator has to decrease for the sum rate to decrease, i.e., $\eta_{1k}>\eta_{1(k+1)}\geq0$. From complementary slackness, we must have $r_{1k}=0$. Therefore, in order for the sum rate to decrease we must have $r_{2k}>r_{2(k+1)}$, which in turn leads to $\eta_{2k}=0$.

From (\ref{eq_bc_weakrate}), we know that for the weak user's rate to decrease, the numerator has to decrease, i.e., we must have $\eta_{2(k+1)}-\eta_{1(k+1)}<\eta_{2k}-\eta_{1k}$. Since $\eta_{2k}=0$, this is equivalent to having $\eta_{2(k+1)}<\eta_{1(k+1)}-\eta_{1k}$. However, we know from above that $\eta_{1k}>\eta_{1(k+1)}$, i.e., $\eta_{2(k+1)}<0$, an obvious contradiction by non-negativity of the Lagrange multipliers.
\end{Proof}

\begin{lemma}\label{thm_bc_weakrate}
The weak user's rate $\{r_{2i}^*\}$ is monotonically increasing.
\end{lemma}

\begin{Proof}
We also prove this by contradiction. Assume that there exists some time slot $k$ such that $r_{2k}>r_{2(k+1)}$. From (\ref{eq_bc_weakrate}), since the denominator cannot increase, the numerator has to decrease for the weak user's rate to decrease, i.e., $\eta_{2(k+1)}-\eta_{1(k+1)}<\eta_{2k}-\eta_{1k}$. Let us consider two different cases. First, assume $\eta_{1k} \geq \eta_{1(k+1)}$. Therefore, we must have $\eta_{2k}>\eta_{2(k+1)}+(\eta_{1k}-\eta_{1(k+1)}) \geq 0$, and thus, by complementary slackness, $r_{2k}=0$, and hence, $r_{2(k+1)}$ cannot be less since it cannot drop below zero. Now, assume $\eta_{1k}<\eta_{1(k+1)}$. In this case, by complementary slackness, $r_{1(k+1)}=0$. By Lemma~\ref{thm_bc_sumrate}, we have $r_{1k}+r_{2k}\leq r_{2(k+1)}$, i.e., $r_{2(k+1)}\geq r_{2k}$, which is a contradiction.
\end{Proof}

With the change of variables: $p_{ti} \triangleq 2^{2(r_{1i}+r_{2i})} - 1$, and $p_{2i} \triangleq 2^{2r_{2i}} - 1$, (\ref{eq_opt_bc}) becomes:
\begin{align}
\max_{{\bf p}_t , {\bf p}_2 } \quad &\mu_1\sum_{i=1}^N g\left(p_{ti}\right) + (\mu_2-\mu_1)\sum_{i=1}^N g\left(p_{2i}\right)  \nonumber \\
\mbox{s.t.} \quad &\sum_{i=1}^k (\sigma^2 -1)p_{2i} + p_{ti} \leq \sum_{i=1}^k E_i, \quad \forall k \nonumber \\
& \sum_{i=1}^k p_{ti} \leq \sum_{i=1}^k \bar{E}_{1i}, \quad \forall k \nonumber \\
& \sum_{i=1}^k p_{2i} \leq \sum_{i=1}^k \bar{E}_{2i}, \quad \forall k \nonumber \\
& p_{ti} \geq p_{2i}, \quad \forall i
\end{align}

We now decompose the above problem into an inner and an outer problem and iterate between them until convergence. First, we fix the value of ${\bf p}_2$, and solve the following inner problem:
\begin{align}\label{eq_bc_inner_prb}
H({\bf p}_2) \triangleq \max_{{\bf p}_t } \quad & \sum_{i=1}^N g(p_{ti}) \nonumber \\
\mbox{s.t.} \quad &\sum_{i=1}^k p_{ti} \leq \sum_{i=1}^k V_i, \quad \forall k \nonumber \\
&p_{ti}\geq p_{2i},\quad \forall i
\end{align}
where the modified energy levels $V_i$ are defined as follows
\begin{align}
&V_i=B_i-B_{i-1}, \nonumber \\
&B_i = \min\left\{ \sum_{j=1}^i \bar{E}_{1j}, \sum_{j=1}^i E_j - (\sigma^2-1)p_{2j} \right\}, \quad B_0=0
\end{align}
We have the following lemma for this inner problem whose proof is similar to that of Lemma~\ref{thm_inner_prb_ccv}.

\begin{lemma}\label{thm_bc_inner_ccv}
$H({\bf p}_2)$ is a decreasing concave function in ${\bf p}_2$.
\end{lemma}

We note that the ${\bf p}_2$ vector serves as a \emph{minimum power constraint} to the inner problem. Let us write the Lagrangian for the inner problem:
\begin{align}
\mathcal{L}=&-\sum_{i=1}^N g(p_{ti}) + \sum_{j=1}^N \lambda_j \left( \sum_{i=1}^j p_{ti} - \sum_{i=1}^j V_i \right) \nonumber
\\&- \sum_{i=1}^N \mu_i \left(p_{ti}-p_{2i}\right)
\end{align}
Taking the derivative with respect to $p_{ti}$ and equating to zero, we obtain:
\begin{align}
p_{ti}=\frac{1}{\sum_{j=i}^N\lambda_j - \mu_i} -1
\end{align}
First, let us examine the necessary conditions for the optimal power to increase, i.e., $p_{ti}<p_{t(i+1)}$. This occurs iff $\lambda_i + \mu_{i+1} > \mu_i \geq0$. Thus, we must either have $\lambda_i > 0$ which means that, by the complementary slackness, we have to consume all the available energy by the end of the $i$th slot. Or, we have $\mu_{i+1}>0$ which means that $p_{t(i+1)}=p_{2(i+1)}$. Next, let us examine the necessary conditions for the optimal power to decrease, i.e., $p_{ti}>p_{t(i+1)}$. This occurs iff $\mu_i > \lambda_i + \mu_{i+1} \geq 0$, and therefore, we must have $p_{ti}=p_{2i}$.

We note from Lemmas~\ref{thm_bc_sumrate} and \ref{thm_bc_weakrate} that both $\{p^*_{2i}\}$ and $\{p^*_{ti}\}$ are monotonically increasing. Therefore, we only focus on fixing an increasing feasible ${\bf p}_2$. This, when combined with the above conditions, leads to the following lemma.
\begin{lemma}
For a fixed increasing ${\bf p}_2$, the optimal solution ${\bf p}_t$ of the inner problem is also increasing.
\end{lemma}

\begin{Proof}
By the KKT conditions stated above, if we have $p_{ti}>p_{t(i+1)}$, then we must have $p_{ti}=p_{2i}$. Thus, we will have $p_{t(i+1)}<p_{ti}=p_{2i}\leq p_{2(i+1)}$, i.e., the minimum power constraint is not satisfied at the $i+1$st slot.
\end{Proof}

Therefore, choosing an increasing ${\bf p}_2$ in the outer problem ensures that the inner problem's solution ${\bf p}_t$ is also increasing, and thereby, satisfies the conditions of Lemmas~\ref{thm_bc_sumrate} and \ref{thm_bc_weakrate}. We solve the inner problem by Algorithm \ref{alg_bc_inner}. The algorithm's main idea is to equalize the powers as much as possible via directional water-filling \cite{omurFade} while satisfying the minimum power requirements.

\begin{algorithm}[t]
\caption{}
\begin{algorithmic}[1] \label{alg_bc_inner}
\STATE Initialize the status of each bin $S_i=V_i$
\STATE Mark bins by their minimum power requirements $\{p_{2i}\}_{i=1}^N$
\STATE Set $k=N$
\WHILE{$k\geq1$}
\IF{$S_k<p_{2k}$}
\STATE Pour water into the $k$th bin from previous bins, in a backward manner, until equality holds
\ELSE
\STATE Do directional water-filling over the current and upcoming bins $\{k,k+1,\dots,N\}$
\ENDIF
\STATE Update the status of each bin
\STATE $k \leftarrow k-1$
\ENDWHILE
\end{algorithmic}
\end{algorithm}

Observe that the algorithm gives a feasible power profile; it examines each slot, and does not move backwards unless the minimum power requirement is satisfied. If there is an excess energy above the minimum, say at slot $k$, it performs directional water-filling which will occur if $S_k>S_{k+1}$ (let us consider water-filling only over two bins for simplicity). Since the minimum power requirement vector ${\bf p}_2$ is increasing, after equalizing the energies the updated status will satisfy $S_k=S_{k+1}>p_{2(k+1)}\geq p_{2k}$, i.e., the minimum power requirement is always satisfied if directional water-filling occurs. Also observe that the algorithm cannot give out a decreasing power profile since ${\bf p}_2$ is increasing.

According to the KKT conditions, the power increases from slot $k$ to slot $k+1$ only if $p_{t(k+1)}=p_{2(k+1)}$ or the total energy is consumed by slot $k$. We see that the algorithm satisfies this condition. Power increases only if directional water-filling is not applied at slot $k$, which means that either some of the water was poured forward in the previous iteration to satisfy $p_{t(k+1)}=p_{2(k+1)}$, or no water was poured which means that all energy is consumed by slot $k$.

\begin{figure*}[t]
\center
\includegraphics[scale=0.8]{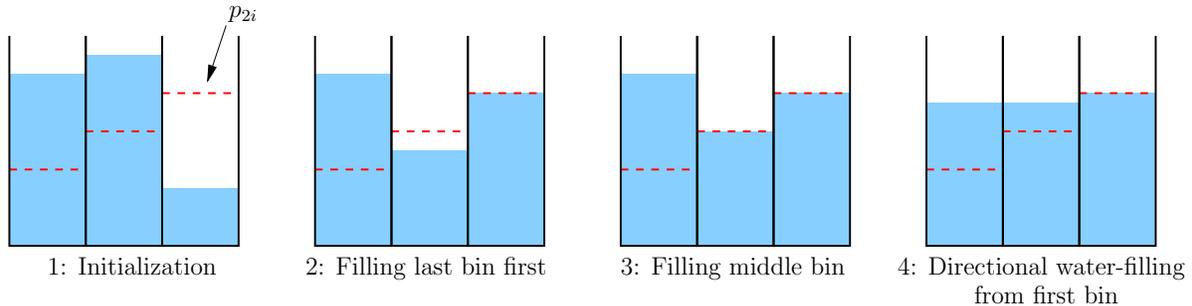}
\caption{Numerical example for the BC inner problem.}
\label{fig_bc_inner}
\end{figure*}

A numerical example for a three-slot system is shown in Fig.~\ref{fig_bc_inner}. The minimum power requirements are shown by red dotted lines in each bin. According to the algorithm, we first initialize by pouring all the amounts of water in their corresponding bins. We begin by checking the last bin, and we see that it needs some extra water to satisfy its minimum power requirement. Thus, we pour water forward from the middle bin until the minimum power requirement of the last bin is satisfied with equality. This causes a deficiency in the middle bin, and therefore, we pour water forward from the first bin until the minimum power requirement of the middle bin is satisfied with equality. Since the problem is feasible, the amount of water remaining in the first bin should satisfy its minimum power requirement. In fact, in this example, there is an excess amount that is therefore used to equalize the water levels of the first two bins via directional water-filling. This ends the algorithm and gives the optimum power profile.

We now find the optimum value of ${\bf p}_2$ by solving the following outer problem:
\begin{align}
\max_{{\bf p}_2 } \quad &\mu H\left({\bf p}_2\right) + \sum_{i=1}^N g(p_{2i}) \nonumber \\
\mbox{s.t.} \quad &\sum_{i=1}^k p_{2i} \leq \sum_{i=1}^k K_i, \quad \forall k
\end{align}
where $\mu \triangleq \frac{\mu_1}{\mu_2-\mu_1}$, and the modified water levels $K_i$ are given by:
\begin{align}
&K_i=A_i-A_{i-1}, \nonumber \\
&A_i = \min\left\{ \sum_{j=1}^i \bar{E}_{2j}, \sum_{j=1}^i \bar{E}_{1j}, \frac{1}{\sigma^2}\sum_{j=1}^i E_j \right\}, \quad A_0=0
\end{align}
where the extra terms in the $A_i$ expression are to ensure feasibility of the inner problem. By Lemma~\ref{thm_bc_inner_ccv}, the outer problem is a convex optimization problem \cite{boyd}. We solve it by an algorithm similar to that of the two-hop network outer problem, except that we only focus on choosing increasing power vectors ${\bf p}_2$ in each iteration. By convexity of the problem, the iterations converge to the optimal solution.

\section{Numerical Results}

In this section, we present numerical results for the considered systems models. We focus on the specific case where $g(x)=\log(1+x)$, and $\phi=g^{-1}$. Starting with the single-user channel, we consider a five-slot system with energy amounts of ${\bf E}=[2,2,1,2.5,0.5]$ at the transmitter, and ${\bf \bar{E}}=[1,1,0.5,2.5,3]$ at the receiver. The optimal rates in this case according to Theorem~\ref{thm_p2p_sol} are given by ${\bf r}^*=[0.6061,0.6061,0.6061,1.2528,1.3863]$. As we see, the rates are non-decreasing, which is consistent with Lemma~\ref{thm_p2p_inc}, and they strictly increase only after consuming all the receiver's energies in decoding by the end of the third slot, and again by the end of the fourth one, which is consistent with Lemma~\ref{thm_p2p_consume}.

\begin{figure}[t]
\center
\includegraphics[scale=0.45]{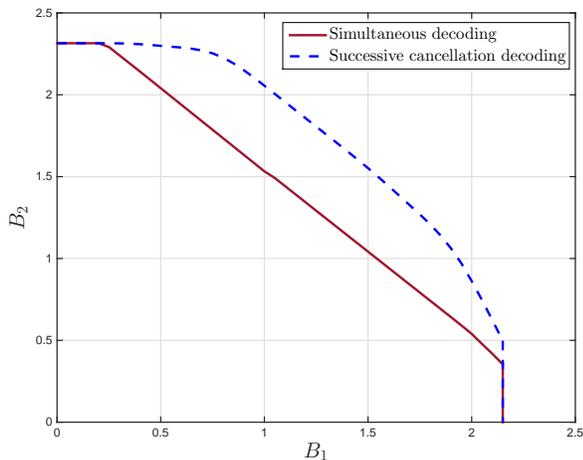}
\caption{Departure regions of a MAC with simultaneous and successive cancellation decoding.}
\label{fig_mac_sim}
\end{figure}

In Fig.~\ref{fig_mac_sim}, we plot the maximum departure regions for a MAC with simultaneous decoding and successive cancellation decoding. We consider a system of three time slots, during which the nodes harvest the energies: ${\bf E}_1=[0.5,1,2]$, ${\bf E}_2=[1,2,0.5]$, and $\bar{{\bf E}}=[1.5,2,0.5]$. We observe that the simultaneous decoding region lies strictly inside the successive decoding region. The latter, given by the geometric programming framework, is only a local optimal solution; one can therefore achieve even higher rates if a global optimal solution is attained.

\begin{figure}[t]
\center
\includegraphics[scale=0.45]{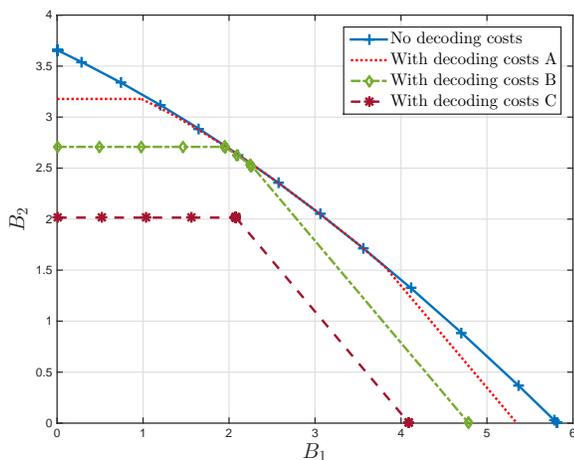}
\caption{Departure regions of a BC with and without decoding costs.}
\label{fig_bc_sim}
\end{figure}

Finally, in Fig.~\ref{fig_bc_sim}, we provide some simulation results to illustrate the difference between the departure regions with and without decoding costs for a BC. We consider a system of three time slots, where the energy profile of the transmitter is given by ${\bf E}=[5,6,7]$. The maximum departure region with no decoding costs is shown in blue. We vary the energy profiles at the receivers to show the effect of the decoding costs on the maximum departure region. We start by setting $\bar{{\bf E}}_1=[4,5,6]$, and $\bar{{\bf E}}_2=[1,2,3]$, to get region $A$ in red. Then we lower the values to $\bar{{\bf E}}_1=[3,4,5]$, and $\bar{{\bf E}}_2=[1,1.5,2]$, to get region $B$ in green. Finally, we lower the values again to $\bar{{\bf E}}_1=[2,3,4]$, and $\bar{{\bf E}}_2=[0.5,1,1.5]$, to get region $C$ in brown. We note that as we lower the energy profiles at the receivers, the decoding causality constraints become more binding, and therefore, the region progressively shrinks.

\section{Conclusions and Future Directions}

We considered decoding costs in energy harvesting communication networks. In our settings, we assumed that receivers, in addition to transmitters, rely on energy harvested from nature. Receivers need to spend a decoding power that is a function of the incoming rate in order to receive their packets. This gave rise to the {\it decoding causality} constraints: receivers cannot spend energy in decoding prior to harvesting it. We first considered a single-user setting and maximized the throughput by a given deadline. Next, we considered two-hop networks and characterized the end-to-end throughput maximizing policies. Then, we considered two-user MAC and BC settings, with focus on exponential decoding functions, and characterized the maximum departure regions. In most of the models considered, we were able to move the receivers' decoding costs effect back to the transmitters as {\it generalized data arrivals}; transmitters need to adapt their powers (and rates) not only to their own energies, but to their intended receivers' energies as well. Such adaptation is governed by the characteristics of the decoding function.

Throughout this paper, we only considered receiver decoding costs in our models without considering transmitter processing costs. On the other hand, other works have considered the processing costs at the transmitter \cite{erkipCost, ruiZhangNonIdeal, orhan-broadband, omurHybrid} without considering decoding costs at the receiver. In their models, the transmitter spends a constant amount of power per unit time whenever it is communicating to account for circuitry processing; while in our model, the receiver spends a decoding power which is a function of the incoming data rate. As a future work, the two approaches can be combined to account for both the processing costs at the transmitter and the decoding costs at the receiver in a single setting.

\begin{thebibliography}{10}

\providecommand{\url}[1]{#1}
\csname url@samestyle\endcsname
\providecommand{\newblock}{\relax}
\providecommand{\bibinfo}[2]{#2}
\providecommand{\BIBentrySTDinterwordspacing}{\spaceskip=0pt\relax}
\providecommand{\BIBentryALTinterwordstretchfactor}{4}
\providecommand{\BIBentryALTinterwordspacing}{\spaceskip=\fontdimen2\font plus
\BIBentryALTinterwordstretchfactor\fontdimen3\font minus
  \fontdimen4\font\relax}
\providecommand{\BIBforeignlanguage}[2]{{%
\expandafter\ifx\csname l@#1\endcsname\relax
\typeout{** WARNING: IEEEtran.bst: No hyphenation pattern has been}%
\typeout{** loaded for the language `#1'. Using the pattern for}%
\typeout{** the default language instead.}%
\else
\language=\csname l@#1\endcsname
\fi
#2}}
\providecommand{\BIBdecl}{\relax}
\BIBdecl

\bibitem{jingP2P}
J.~Yang and S.~Ulukus, ``Optimal packet scheduling in an energy harvesting
  communication system,'' \emph{IEEE Trans. Comm.}, vol.~60, no.~1, pp.
  220--230, January 2012.

\bibitem{kayaEmax}
K.~Tutuncuoglu and A.~Yener, ``Optimum transmission policies for battery
  limited energy harvesting nodes,'' \emph{IEEE Trans. Wireless Comm.},
  vol.~11, no.~3, pp. 1180--1189, March 2012.

\bibitem{omurFade}
O.~Ozel, K.~Tutuncuoglu, J.~Yang, S.~Ulukus, and A.~Yener, ``Transmission with
  energy harvesting nodes in fading wireless channels: Optimal policies,''
  \emph{IEEE JSAC}, vol.~29, no.~8, pp. 1732--1743, September 2011.

\bibitem{ruiZhangEH}
C.~K. Ho and R.~Zhang, ``Optimal energy allocation for wireless communications
  with energy harvesting constraints,'' \emph{IEEE Trans. Signal Proc.},
  vol.~60, no.~9, pp. 4808--4818, September 2012.

\bibitem{jingMAC}
J.~Yang and S.~Ulukus, ``Optimal packet scheduling in a multiple access channel
  with energy harvesting transmitters,'' \emph{Journal of Communications and
  Networks}, vol.~14, no.~2, pp. 140--150, April 2012.

\bibitem{jingBC}
J.~Yang, O.~Ozel, and S.~Ulukus, ``Broadcasting with an energy harvesting
  rechargeable transmitter,'' \emph{IEEE Transactions on Wireless
  Communications}, vol.~11, no.~2, pp. 571--583, February 2012.

\bibitem{elifBC}
M.~A. Antepli, E.~Uysal-Biyikoglu, and H.~Erkal, ``Optimal packet scheduling on
  an energy harvesting broadcast link,'' \emph{IEEE Journal on Selected Areas
  in Communications}, vol.~29, no.~8, pp. 1721--1731, September 2011.

\bibitem{elifDataBC}
H.~Erkal, F.~M. Ozcelik, and E.~Uysal-Biyikoglu, ``Optimal offline broadcast
  scheduling with an energy harvesting transmitter,'' \emph{EURASIP Journal on
  Wireless Communications and Networking}, vol. 2013, no.~1, pp. 1--20, July
  2013.

\bibitem{omurBC}
O.~Ozel, J.~Yang, and S.~Ulukus, ``Optimal broadcast scheduling for an energy
  harvesting rechargebale transmitter with a finite capacity battery,''
  \emph{IEEE Transactions on Wireless Communications}, vol.~11, no.~6, pp.
  2193--2203, June 2012.

\bibitem{gunduz2Hop}
D.~Gunduz and B.~Devillers, ``Two-hop communication with energy harvesting,''
  in \emph{IEEE CAMSAP}, December 2011.

\bibitem{ruiZhangRelay}
C.~Huang, R.~Zhang, and S.~Cui, ``Throughput maximization for the {G}aussian
  relay channel with energy harvesting constraints,'' \emph{IEEE Journal on
  Selected Areas in Communications}, vol.~31, no.~8, pp. 1469--1479, August
  2013.

\bibitem{erkipRelay}
O.~Orhan and E.~Erkip, ``Optimal transmission policies for energy harvesting
  two-hop networks,'' in \emph{CISS}, March 2012.

\bibitem{schoberRelay}
I.~Ahmed, A.~Ikhlef, R.~Schober, and R.~K. Mallik, ``Power allocation in energy
  harvesting relay systems,'' in \emph{IEEE VTC}, May 2012.

\bibitem{letaiefRelay}
Y.~Luo, J.~Zhang, and K.~B. Letaief, ``Optimal scheduling and power allocation
  for two-hop energy harvesting communication systems,'' \emph{IEEE
  Transactions on Wireless Communications}, vol.~12, no.~9, pp. 4729--4741,
  September 2013.

\bibitem{varanBuffer}
B.~Varan and A.~Yener, ``Two-hop networks with energy harvesting: The
  (non-)impact of buffer size,'' in \emph{IEEE GlobalSIP}, December 2013.

\bibitem{berkCoop}
B.~Gurakan, O.~Ozel, J.~Yang, and S.~Ulukus, ``Energy cooperation in energy
  harvesting communications,'' \emph{IEEE Transactions on Communications},
  vol.~61, no.~12, pp. 4884--4898, December 2013.

\bibitem{berkDiamond}
B.~Gurakan and S.~Ulukus, ``Energy harvesting diamond channel with energy
  cooperation,'' in \emph{IEEE ISIT}, July 2014.

\bibitem{kayaLoss}
K.~Tutuncuoglu, A.~Yener, and S.~Ulukus, ``Optimum policies for an energy
  harvesting transmitter under energy storage losses,'' \emph{IEEE Journal on
  Selected Areas in Communications}, vol.~33, no.~3, pp. 476--481, March 2015.

\bibitem{gunduzLoss}
D.~Gunduz and B.~Devillers, ``A general ramework for the optimization of energy
  harvesting communication systems with battery imperfections,'' \emph{Journal
  of Communications and Networks}, vol.~14, no.~2, pp. 130--139, April 2012.

\bibitem{erkipCost}
O.~Orhan, D.~Gunduz, and E.~Erkip, ``Throughput maximization for an energy
  harvesting communication system with processing cost,'' in \emph{IEEE ITW},
  September 2012.

\bibitem{ruiZhangNonIdeal}
J.~Xu and R.~Zhang, ``Throughput optimal policies for energy harvesting
  wireless transmitters with non-ideal circuit power,'' \emph{IEEE JSAC},
  vol.~32, no.~2, pp. 322--332, February 2014.

\bibitem{orhan-broadband}
O.~Orhan, D.~Gunduz, and E.~Erkip, ``Energy harvesting broadband communication
  systems with processing energy cost,'' \emph{IEEE Trans. Wireless Comm.},
  vol.~13, no.~11, pp. 6095--6107, November 2014.

\bibitem{omurHybrid}
O.~Ozel, K.~Shahzad, and S.~Ulukus, ``Optimal energy allocation for energy
  harvesting transmitters with hybrid energy storage and processing cost,''
  \emph{IEEE Trans. Signal Proc.}, vol.~62, no.~12, pp. 3232--3245, June 2014.

\bibitem{payaroMercury}
M.~Gregori and M.~Payar{\'o}, ``Optimal power allocation for a wireless
  mutli-antenna energy harvesting node with arbitrary input distribution,'' in
  \emph{IEEE ICC}, June 2012.

\bibitem{letaiefTraining}
Y.~Luo, J.~Zhang, and K.~B. Letaief, ``Training optimization for energy
  harvesting communication systems,'' in \emph{IEEE Globecom}, December 2012.

\bibitem{vvvEH}
A.~Nayyar, T.~Basar, D.~Teneketzis, and V.~V. Veeravalli, ``Optimal strategies
  for communication and remote estimation with an energy harvesting sensor,''
  \emph{IEEE Trans. on Automatic Control}, vol.~58, no.~9, pp. 2246--2260,
  September 2013.

\bibitem{ruiZhangOutage}
C.~Huang, R.~Zhang, and S.~Cui, ``Optimal power allocation for outage
  probability minimization in fading channels with energy harvesting
  constraints,'' \emph{IEEE Transactions on Wireless Communications}, vol.~13,
  no.~2, pp. 1074--1087, February 2014.

\bibitem{varan-continuous}
B.~Varan, K.~Tutuncuoglu, and A.~Yener, ``Energy harvesting communications with
  continuous energy arrivals,'' in \emph{UCSD ITA}, Feb 2014.

\bibitem{erkipDelay}
O.~Orhan, D.~Gunduz, and E.~Erkip, ``Delay-constrained distortion minimization
  for energy harvesting transmission over a fading channel,'' in \emph{IEEE
  ISIT}, July 2013.

\bibitem{aggarwalPmax}
Z.~Wang, V.~Aggarwal, and X.~Wang, ``Iterative dynamic water-filling for fading
  multiple-access channels with energy harvesting,'' \emph{IEEE Journal on
  Selected Areas in Communications}, vol.~33, no.~3, pp. 382--395, March 2015.

\bibitem{kayaRxEH}
K.~Tutuncuoglu and A.~Yener, ``Communicating with energy harvesting
  transmitters and receivers,'' in \emph{UCSD ITA}, February 2012.

\bibitem{yatesRxEH1}
H.~Mahdavi-Doost and R.~D. Yates, ``Energy harvesting receivers: Finite battery
  capacity,'' in \emph{IEEE ISIT}, July 2013.

\bibitem{yatesRxEH2}
R.~D. Yates and H.~Mahdavi-Doost, ``Energy harvesting receivers: Optimal
  sampling and decoding policies,'' in \emph{IEEE GlobalSIP}, December 2013.

\bibitem{yatesRxEH3}
H.~Mahdavi-Doost and R.~D. Yates, ``Fading channels in energy-harvesting
  receivers,'' in \emph{CISS}, March 2014.

\bibitem{grover}
P.~Grover, K.~Woyach, and A.~Sahai, ``Towards a communication-theoretic
  understanding of system-level power consumption,'' \emph{IEEE Journal on
  Selected Areas in Communications}, vol.~29, no.~8, pp. 1744--1755, September
  2011.

\bibitem{payaroRxEH}
J.~Rubio, A.~Pascual-Iserte, and M.~Payar{\'o}, ``Energy-efficient resource
  allocation techniques for battery management with energy harvesting nodes: a
  practical approach,'' in \emph{European Wireless Conference}, April 2013.

\bibitem{rost-decPower}
P.~Rost and G.~Fettweis, ``On the transmission-computation-energy tradeoff in
  wireless and fixed networks,'' in \emph{IEEE Globecom Workshop on Green
  Communications}, December 2010.

\bibitem{goldsmith-decPower}
S.~Cui, A.~J. Goldsmith, and A.~Bahai, ``Power estimation for {V}iterbi
  decoders,'' Wireless System Lab, Stanford Univ., CA, 2003, available Online:
  \url{http://wsl.stanford.edu/publications.html}.

\bibitem{elgamalKim}
A.~E. Gamal and Y.~Kim, \emph{Network Information Theory}.\hskip 1em plus 0.5em
  minus 0.4em\relax Cambridge University Press, 2011.

\bibitem{succCVXapprox}
B.~R. Marks and G.~P. Wright, ``A general inner approximation algorithm for
  nonconvex mathematical programs,'' \emph{Operations research}, vol.~26,
  no.~4, pp. 681--683, July-August 1978.

\bibitem{palomarGP}
M.~Chiang, C.~W. Tan, D.~P. Palomar, D.~O'Neill, and D.~Julian, ``Power control
  by geometric programming,'' \emph{IEEE Transactions on Wireless
  Communications}, vol.~6, no.~7, pp. 2640--2651, July 2007.

\bibitem{cover}
T.~Cover and J.~A. Thomas, \emph{Elements of Information Theory}, 2006.

\bibitem{boyd}
S.~P. Boyd and L.~Vandenberghe, \emph{Convex Optimization}, 2004.

\bibitem{palomarCVXsep}
A.~A. D'Amico, L.~Sanguinetti, and D.~P. Palomar, ``Convex seperable problems
  with linear constraints in signal processing and communications,'' \emph{IEEE
  Transactions on Signal Processing}, vol.~62, no.~22, pp. 6045--6058, November
  2014.

\bibitem{arafaGlobalSIP}
A.~Arafa and S.~Ulukus, ``Single-user and multiple-access channels with energy
  harvesting transmitters and receivers,'' in \emph{IEEE GlobalSIP}, December
  2014.

\end{thebibliography}
\end{document}